\documentclass[twocolumn]{aa}
\usepackage{graphicx}
\usepackage[]{natbib}
\bibpunct{(}{)}{;}{a}{}{,} 
\usepackage{txfonts}
\newcommand{\al}{\alpha}
\newcommand{\lam}{\lambda}
\newcommand{\mum}{\mbox{$\,\mu \mathrm{m}$}}
\newcommand{\Brd}{\mbox{$\mathrm{Br\delta}$}}

\newcommand{\Brg}{\mbox{$\mathrm{Br\gamma}$}}
\newcommand{\Fe}{\mbox{$\mathrm{[FeII]}$}}
\newcommand{\Fzwei}{\mbox{$\mathrm{[FeII]_{ 1.257}}$}}
\newcommand{\Fsechs}{\mbox{$\mathrm{[FeII]_{ 1.644}}$}}
\newcommand{\Ha}{\mbox{$\mathrm{H\al}$}}
\newcommand{\Hb}{\mbox{$\mathrm{H\beta}$}}
\newcommand{\Hzwei}{\mbox{$\mathrm{H_2}$}}
\newcommand{\Hzweiseins}{\mbox{$\mathrm{H_2}\:v\!=\!1\!-\!0\:\mathrm{S(1)}$}}
\newcommand{\Hzweisdrei}{\mbox{$\mathrm{H_2}\:v\!=\!1\!-\!0\:\mathrm{S(3)}$}}
\newcommand{\Oi}{\mbox{$\mathrm{[OI]_{6300}}$}}
\newcommand{\OI}{\mbox{$\mathrm{[OI]}$}}
\newcommand{\Pa}{\mbox{$\mathrm{Pa\al}$}}
\newcommand{\Pb}{\mbox{$\mathrm{Pa\beta}$}}
\newcommand{\Si}{\mbox{$\mathrm{[SiVI]}$}}
\newcommand{\Ho}{\mbox{$\mathrm{H_{o}}$}}
\newcommand{\Mpc}{\mbox{$\,\mathrm{Mpc}$}}
\newcommand{\kms}{\mbox{$\,\mathrm{km}\,\mathrm{s^{-1}}$}}
\newcommand{\Lir}{\,L_{\mathrm{IR}}}
\newcommand{\LPa}{\,L_{\mathrm{Pa\al}}}
\newcommand{\qo}{\mbox{$\mathrm{q_{o}}$}}
\begin{document}
   \title{Follow-Up Near-infrared Spectroscopy of Ultraluminous Infrared Galaxies
     observed by ISO\thanks{Based on observations collected at the
European Southern Observatory, Chile, ESO No. 62.P-0315(A) and 63.P-0264(A).}}

   \subtitle{}

   \author{H. Dannerbauer \inst{1,2}\thanks{Present address.}
          \and
           D. Rigopoulou \inst{1,3}$^{\star\star}$
	  \and 
	   D. Lutz \inst{1}
	  \and 
	   R. Genzel \inst{1,4}
	  \and 
	   E. Sturm \inst{1}
	  \and
	   A.F.M. Moorwood \inst{5}
          }

   \offprints{H. Dannerbauer,
email: {dannerb@mpia-hd.mpg.de}}

   \institute{
              Max-Planck-Institut f\"ur extraterrestrische Physik, 
              Postfach 1312, 85741 Garching, Germany\\
              \email{lutz@mpe.mpg.de,genzel@mpe.mpg.de,sturm@mpe.mpg.de}
	      \and
	      Max-Planck-Institut f\"ur Astronomie,
              K\"onigstuhl 17,  69117 Heidelberg, Germany \\
	      \email{dannerb@mpia-hd.mpg.de} 
              \and
	      Astrophysics, Department of Physics, Keble Road, Oxford OX1
              3 RH, U.K. \\
             \email{dar@astro.ox.ac.uk}
	      \and
	      Department of Physics, 366 Le~Conte Hall, University of 
	      California, Berkeley, CA 94720-7300	      
	      \and
	      European Southern Observatory, Karl-Schwarzschild-Str. 2,
              85748 Garching, Germany\\
              \email{amoor@eso.org}
             }

   \date{Received date; accepted}

   \abstract{We present low resolution near-infrared spectroscopy of
an unbiased sample of 24 ultraluminous infrared galaxies (ULIRGs),
selected from samples previously observed spectroscopically in the
mid-infrared with the Infrared Space Observatory (ISO). Qualitatively,
the near-infrared spectra resemble those of starbursts.  Only in one
ULIRG, IRAS~04114$-$5117E, do we find spectroscopic evidence for AGN
activity. The spectroscopic classification in the near-infrared is in
very good agreement with the mid-infrared one.  For a subset of our
sample for which extinction corrections can be derived from $\Pa$ and
$\Brg$, we find rather high $\Pa$ luminosities, in accordance with the
powering source of these galaxies being star formation.  $\Fe$
emission is strong in ULIRGs and may be linked to starburst and
superwind activity. Additionally, our sample includes two unusual
objects. The first, IRAS~F00183$-$7111, exhibits extreme $\Fe$
emission and the second, IRAS~F23578$-$5307, is according to our knowledge one of the
most luminous infrared galaxies in $\Hzwei$ rotation-vibration
emission.
   }
\titlerunning{NIR-Spectroscopy of ULIRGs observed by ISO}
\authorrunning{H. Dannerbauer et al.}
   \maketitle
%

\section{Introduction}
\label{introduction}
  Ultraluminous infrared galaxies \citep[see][for a review]{san96}
radiate the bulk of their energy in the far-infrared as a continuum
emitted by warm dust ($T \sim 40 - 60$ K). The dust is heated by
ultraviolet radiation either coming from an immense starburst in the
central region of these objects or from an accretion disk surrounding
a black hole. Since their discovery by IRAS there has been an intense
debate about the nature of ULIRGs and their possible evolutionary
connection to quasars.  \citet{san88a} postulated that the dominant
energy source in a ULIRG is a dust enshrouded quasar appearing in the
last phase of the merging of two or more gas-rich disk
galaxies. Although evidence for both starburst and AGN activity in
ULIRGs has been found, the question of what generally dominates the
luminosity has remained largely unresolved, mainly due to
observational difficulties associated with the large dust
obscuration. Since extinction in the optical is about a factor of 10
higher than in the near-infrared (e.g., \citealt{mat90}) and as the
latter provides both starburst signatures and direct spectroscopic
indicators of AGN activity (e.g., \citealt{mar94}), near-infrared
spectroscopy is an important diagnostic for these objects.  With the
advent of sensitive near-infrared detectors, near-infrared
spectroscopy of ultraluminous galaxies
\citep[e.g.,][]{gol95,gol97a,gol97b,vei97b,vei99b,mur99,mur00,mur01,bur01,val05}
began to unveil the nature of these most luminous galaxies in the
local universe. Most near-infrared spectra of ULIRGs were found to be
starburst-like, but the starburst luminosity traced by the
observations appeared typically insufficient to account for the full
ULIRG luminosities.

With ISO \citep{kes96} it became possible to study even more obscured
regions by means of sensitive mid-infrared spectroscopy. Fine
structure lines in 15 ULIRGs \citep{gen98} were observed with the ISO
short-wavelength grating spectrometer.  By using the
spectrophotometric mode of the photometer on board ISO, low resolution
spectroscopy of the mid-infrared aromatic features (`PAHs') was
obtained for a sample of 62 ULIRGs (\citealt{lut98,rig99}).  Finally,
16 ULIRGs of particularly high luminosity were observed by
\cite{tra01} with the low resolution spectroscopic mode of the ISOCAM
camera.  These studies of 78 ULIRGs, drawn mainly from the complete
IRAS 2 Jy sample \citep{str90,str92} and its extension, the IRAS 1.2
Jy sample \citep{fis95}, find that $70-80$~\% of the ULIRGs are
dominated by starburst activity but that there is also a trend towards
AGN dominance at the highest luminosities. \citet{lut99} compared ISO
classifications (starburst or AGN) with classifications from optical
spectroscopy and found a good agreement if optical LINER spectra were
assigned to the starburst group, supporting a scenario in which
infrared-selected LINERs are due to shocks that are probably related
to galactic superwinds.

We carried out follow-up low resolution near-infrared spectroscopy for
a subset of the ISO ULIRG sample, accessible from La Silla, in order
to search for obscured AGNs and to compare the ISO indicators with
those from the near-infrared wavelength regime.  We used the $\Pa$
line which dominates the near-infrared spectra of ULIRGs and the
coronal $\Si$ line as diagnostic tools to search for broad-line
components and high energy photons, respectively.  We investigate the
so-called `$\Brg$-deficit' quantifying the starburst contribution to
the total luminosity and first studied by \citet{gol95,gol97a,gol97b},
and discuss the strong $\Fe$ emission in ULIRGs and its possible
implications for the nature of infrared-selected LINERs. Finally, we
discuss two unusual near-infrared ULIRG spectra.

The structure of this paper is as follows: In section (2) we present
the sample. In section (3) we report on the observations and data
reduction, followed by section (4), discussing the results of our
near-infrared spectroscopy of ULIRGs. Section (5) summarizes our main
conclusions.  We adopt $\Ho=75\kms$/$\Mpc$ and $\qo = 0.5$.

\section{The Sample}
\label{Sample}
The unbiased ISO sample, comprising observations made with ISO-SWS
\citep{gen98}, ISOPHOT-S \citep{lut98,rig99} and ISOCAM-CVF
\citep{tra01}, was drawn from the complete IRAS 2 Jy sample
\citep{str90,str92} and its extension, the IRAS 1.2 Jy sample
\citep{fis95}. The selected objects had to fulfill the following
criteria: $L_{40-120~\mum}>10^{11.7}$ L$_{\sun}$, equivalent to
$\Lir(8-1000~\mum)>10^{12}$ L$_{\sun}$\footnote{ \citet{per87}
calculates the infrared luminosity by using the four IRAS-bands:
$L(8-1000~\mu m)=4\pi D^{2}[1.8\times
10^{-14}(13.48f_{12}+5.16f_{25}+2.58f_{60}+f_{100})][W]$.}, flux
S$_{60}>\ 1.3$~Jy, good ISO visibility and (for ISOPHOT-S only)
redshift below 0.3.  The sample of 24 ULIRGs presented here is a
subset of targets accessible for observations from La Silla during our
observing runs. We did not bias the sample towards starbursts or
AGNs. Our ULIRGs cover a redshift range $0.043-0.327$.  $\Pa$ is
shifted into the K band for all objects, except for IRAS~F00183$-$7111
which shows $\Pb$ shifted into the H band. The sample and its basic
properties are listed in Table~\ref{tab:sample}.

\section{Observations and Data Reduction}
\label{Observations and Data Reduction}
Near-infrared long-slit spectra of the 24 ULIRGs were obtained with
 SOFI \citep{mor98}, the camera$/$spectrometer on the ESO NTT 3.5~m
 telescope at La Silla/Chile, during two runs in November 1998 and
 August 1999. We chose the red grism covering a spectral range from
 $1.53-2.52~\mum$, including the AGN-indicators $\Pa$ and $\Si$ in
 the K band as well as other diagnostic lines. We used a
 1\arcsec$\times$290\arcsec\/ long slit giving a spectral resolving
 power R ($\equiv \lam / \Delta \lam$) $\sim 600$ at 2~$\mum$. The
 detector used is a Rockwell HgCdTe 1024$\times$1024 Hawaii array with 18.5
 micron pixels and the pixel scale along the spatial axis is 0.292
 arcseconds per pixel.

In single component systems, the slit was oriented along the presumed
major axis of the ULIRG, as determined from K-band images taken by
\citet{rig99} and \citet{bor00}. In addition, a SOFI K-band image was
taken of IRAS~19458$+$0944 which is presented in the appendix of this
paper. In double/multiple component systems, we oriented the slit to
get spectra of two components simultaneously. Table~\ref{tab:sample}
lists the slit position angles used.  Spectra were taken in the
standard way, nodding the telescope $\sim40$\arcsec along the
slit. The total integration time was between 20 to 40 minutes.  Sky
conditions during both runs were photometric with a seeing of
$0.45-0.8$\arcsec\/ and $0.8-1.3$\arcsec, respectively.  Data
reduction was performed using IRAF\footnote{ IRAF is distributed by
the National Optical Astronomy Observatories, which are operated by
the Association of Universities for Research in Astronomy, Inc., under
cooperative agreement with the National Science Foundation.  } and
ECLIPSE software \citep{dev97} in the standard way.  Line parameters
(flux, center, width, see Tables~\ref{tab:fluxes} to \ref{tab:hw})
were measured by fitting a gaussian profile. We estimate flux errors
less than 25\% \footnote{A few ULIRGs of this sample were observed by
\citet{val05} with SOFI in medium resolution mode in the K band. We
note for some line measurements discrepancies. Possible reasons for
that are differences in set-up like position angle and weather
conditions.}.

14 of the 24 ULIRGs contain two or more components (see also
Table~\ref{tab:sample}). If the separation of the components was
sufficient, we extracted spectra for each component, finally obtaining
34 spectra.

\begin{table*}
\caption{Sample characteristics\label{tab:sample}}
\begin{tabular}{lllcrllllr}
\hline\hline
Object&$\Lir$&z&Morpho-&f$_{25}/$f$_{60}$&optical&ISO&RA(2000)&DEC(2000)&PA\\
&&&logy&&type&type&&&degree\\
(1)&(2)&(3)&(4)&(5)&(6)&(7)&(8)&(9)&(10)\\
\hline
F00183$-$7111 &12.77&0.327&S&0.11&LINER$^{1}$&AGN&00 20 34&$-$70 55 26&90.0\\
00188$-$0856 &12.31&0.129&D&0.14&LINER$^{3}$&SB&00 21 26&$-$08 39 27&0.0\\
00199$-$7426 &12.25&0.096&D&0.08&?$^{2}$ &? & 00 22 08&$-$74 09 26&$-$25.5\\
01003$-$2238 &12.19&0.118&S&0.29&HII$^{3}$&SB&01 02 49&$-$22 51 57&0.0\\
01166$-$0844 &11.99&0.118&D&0.10&HII$^{3}$&?&01 19 07&$-$08 29 11&$-$65.0\\
01298$-$0744 &12.28&0.136&S&$<$0.11&HII$^{3}$&SB&01 32 21&$-$07 29 08&0.0\\
01388$-$4618 &12.03&0.090&S&0.12&HII$^{6}$&SB&01 40 55&$-$46 02 53&0.0\\
02411$+$0354 &12.15&0.144&T&0.16&HII$^{3}$&SB&02 43 46&$+$04 06 36&28.0\\
03521$+$0028 &12.46&0.152&D&0.09&LINER$^{3}$&SB&03 54 42&$+$00 37 03&80.6\\
04063$-$3236 &11.97&0.110&D&0.08&&SB&04 08 18&$-$32 28 31&69.0\\
04114$-$5117 &12.16&0.124&D&0.03&&AGN&04 12 44&$-$51 09 40&40.0\\
04384$-$4848 &12.32&0.213&S&0.07&&SB&04 39 50&$-$48 43 16&0.0\\
06009$-$7716 &11.95&0.117&D&0.08&&SB&05 58 37&$-$77 16 39&103.5\\
06035$-$7102 &12.14&0.079&D&0.11&HII$^{2}$ $^{\dagger}$&SB&06 02 54&$-$71 03 09&245.2\\
19458$+$0944 &12.28&0.100&S&$<$0.07&&SB&19 48 15&$+$09 52 01&0.0\\
20049$-$7210 &11.93&0.126&S&0.08&HII$^{4}$&SB&20 10 27&$-$72 01 11&$-$13.0\\
20100$-$4156 &12.55&0.129&D&0.07&HII$^{2}$ $^{\dagger}$&SB&20 13 29&$-$41 47 34&41.1\\
20446$-$6218 &12.04&0.107&D&0.08&HII$^{4}$&?&20 48 44&$-$62 07 25&74.8\\
20551$-$4250 &11.98&0.043&S&0.15&HII$^{2}$ $^{\dagger}$&SB&20 58 26&$-$42 38 57&0.0\\
23128$-$5919 &11.96&0.045&D&0.15&HII$^{2}$&SB&23 15 46&$-$59 03 14&$-$3.0\\
23230$-$6926 &12.22&0.106&D&0.08&LINER$^{2}$&?&23 26 03&$-$69 10 17&22.6\\
23253$-$5415 &12.26&0.129&S&0.09&HII$^{5}$&SB&23 28 06&$-$53 58 31&$-$13.0\\
23389$-$6139 &12.10&0.093&D&0.07&HII$^{2}$ $^{\dagger}$&SB&23 41 43&$-$61 22 50&$-$8.0\\
F23578$-$5307&12.11&0.125 &S &$<$0.05& &? & 00 00 24&$-$52 50 30&90.0\\
\hline
\end{tabular}
\\
Col. (1) --- Name of Object (IRAS ...).\\
Col. (2) --- Log of Infrared Luminosity $\Lir$ in L$_{\sun}$.\\
Col. (3) --- Redshift z.\\
Col. (4) --- S=single; D=double; T=triple.\\
Col. (5) --- Ratio of IRAS flux at 25 and 60~$\mu m$. \\
Col. (6) --- Optical type.\\
Col. (7) --- ISO type; classification based on the PAH feature \citep{rig99,lut99,tra01}.\\
Col. (8) --- RA(J2000).\\
Col. (9) --- DEC(J2000).\\
Col. (10) --- Position Angle PA of the observed ULIRGs.\\
$^{1}$\citet{arm89}; $^{2}$\citet{duc97};
$^{3}$\citet{vei99a}; $^{4}$\citet{all91}; $^{5}$\citet{sek93}; $^{6}$\citet{kew01}.
\\$^{\dagger}$ These ULIRGs were reclassified by \citet{lut99}.
\\?: Quality of these spectra was not good enough for a classification.
\end{table*}
\begin{table*}
\caption{Emission-line fluxes\label{tab:fluxes}}
\tiny
\begin{tabular}{llllllllllllll}
\hline\hline
Object&[FeII]&$\Pb$&[FeII]&$\Hzwei$&$\Pa$&$\Hzwei$&$\Brd$&$\Hzwei+\Si$&$\Hzwei$&HeI&$\Hzwei$&$\Brg$&$\Hzwei$\\
&1.257&1.282&1.644&1.835&1.876&1.891&1.945&1.957&2.033&
2.058&2.122&2.166&2.223\\
(1)&(2)&(3)&(4)&(5)&(6)&(7)&(8)&(9)&(10)&(11)&(12)&
(13)&(14)\\
\hline
F00183$-$7111&0.228:&0.115:&0.376:&&\\
00188$-$0856N&&&&0.079:&0.71&0.039::&&0.123&&&0.104::\\
00188$-$0856S\\
00199$-$7426N&&&&&2.07&0.218&0.066::&0.251:&0.115::&&0.462:\\
00199$-$7426S&&&&&0.126:&&&&&\\
01003$-$2238&&&&&0.991&&&0.190:&\\
01166$-$0844S\\
01166$-$0844W&&&&&0.075&&&\\
01298$-$0744&&&&&0.072&0.006:&&0.017&&&0.025:\\
01388$-$4618&&&&0.12:&1.495&0.083::&&0.220&&0.040::&0.203&0.182::&\\
02411$+$0354Main&&&&&2.906&&0.188&0.099&&0.213:&0.125:&0.408:&\\
02411$+$0354W&&&&&0.680&&0.063:&0.039::&&\\
03521$+$0028&&&&0.162&0.764&&&0.170:&&\\
04063$-$3236NE&&&&0.087:&1.08&&0.055::&0.134&&&0.123:&0.104:&\\
04063$-$3236SW&&&&&0.207&&&&&&&&\\
04114$-$5117E&&&&&0.398&&&0.089:&0.053::&&0.057::&&\\
04114$-$5117SW&&&&&0.082&&&&&&&&\\
04384$-$4848&&&&&1.01&&&&&&&&\\
06009$-$7716E&&&&&0.128&&&&&&&\\
06009$-$7716W&&&&&1.206&&0.102&0.052&&&&0.170::&\\
06035$-$7102W&&&0.232:&&1.122&0.078::&&0.428&0.227&&0.480&\\
06035$-$7102E&&&&&0.231&&&0.516:&&\\
19458$+$0944&&&2.079:&0.911::&8.855&&0.460:&1.276&0.631:&0.484:&1.276:&1.166:&\\
20049$-$7210&&&&&0.172&&\\
20100$-$4156$^{\ast}$&&&&&0.445&&&&&\\
20446$-$6128N&&&&0.050&1.616&0.040:&0.099&0.117&0.025&0.899&0.092:&0.206&\\
20446$-$6128S\\
20551$-$4250&&&0.257&&1.606!&0.011!&0.128!&0.428!&&0.187&0.465&0.199&0.149\\
23128$-$5919N&&&0.960&&13.06!&&0.695!&0.441!&&0.802&0.318&1.191&0.417\\
23128$-$5919S&&&2.038&&26.33!&&1.578!&2.696!&0.451&1.547&1.306&2.675&0.651:\\
23230$-$6926&&&&&0.221&1.561&&0.138&0.278&0.188&0.092&0.247&0.190:\\
23253$-$5415&&&&0.075&0.269&&&0.101&&&0.175&\\
23389$-$6139&&&0.600&0.708:&2.232&0.091&0.121:&0.653&0.187:&&0.581&0.212&0.341::\\
F23578$-$5307&&&&1.395&2.768&0.423::&&1.65&1.136&0.966:&3.735\\
\hline
\end{tabular}
\\
Col. (1) --- Name of Object.\\Col. (2) - Col. (14)
--- Line flux in  $10^{-14}$erg s$^{-1}$ cm$^{-2}$.\\
! --- Line in region of poor atmospheric transmission  between H and
K band.\\
\end{table*}
\begin{table*}
\caption{Observed full width at half maximum of emission lines\label{tab:hw}}
\tiny
\begin{tabular}{llllllllllllll}
\hline\hline
Object&[FeII]&$\Pb$&[FeII]&$\Hzwei$&$\Pa$&$\Hzwei$&$\Brd$&
$\Hzwei+\Si$&$\Hzwei$&HeI&$\Hzwei$&$\Brg$&$\Hzwei$\\
&1.257&1.282&1.644&1.835&1.876&1.891&1.945&1.957&2.033&
2.058&2.122&2.166&2.223\\
(1)&(2)&(3)&(4)&(5)&(6)&(7)&(8)&(9)&(10)&(11)&(12)&
(13)&(14)\\
\hline
F00183$-$7111&1000::&600::&1100::&&&&&&\\
00188$-$0856N&&&&572.6:&481.2&526.7::&&509.4&\\
00188$-$0856S&&&&&\\
00199$-$7426N&&&&&585.3&717.9&&599.6:&417.3::&&823.2:\\
00199$-$7426S&&&&&513.7:&\\
01003$-$2238&&&&&542.1&&&667.6:&&\\
01166$-$0844S&&&&\\
01166$-$0844W&&&&&575.0&&&\\
01298$-$0744&&&&&634.4&&&527.8&&&532.5:&\\
01388$-$4618&&&&682.6:&540.6&656::&&607.2&&&425.0&&\\
02411$+$0354Main&&&&&519.3&&494.9&846.9&&506.3:&401.3::&385:&\\
02411$+$0354W&&&&&468.2&&461.8:&968.0::\\
03521$+$0028&&&&433.7&663.9&&&577.5:&\\
04063$-$3236NE&&&&769.6&587.0&&&500.2&&&377.9::&&\\
04063$-$3236SW&&&&&630&&&\\
04114$-$5117E&&&&&569.8&&&945.9:&&&634.7&\\
04114$-$5117SW&&&&&853.7&&\\
04384$-$4848&&&&&630.7&&&&&\\
06009$-$7716E&&&&&573.8&&&\\
06009$-$7716W&&&&&558.2&&598.5&598.5&&&&381.0::&\\
06035$-$7102W&&&600.1:&&706.6&565.9::&&684.2:&866.3&&794.1\\
06035$-$7102E&&&&&592.9&&&380.8::&&\\
19458$+$0944&&&982.9:&542.4::&524.5&&458.6:&613.6&730.1:&407.2:&561.4:&578.7:&\\
20049$-$7210&&&&&585.2&&&&&\\
20100$-$4156&&&&&892.0&&&&&\\
20446$-$6128N&&&&484.4&502.1&581.7:&461.8&480.0&551.7&557.5&508.8:&643.4&\\
20446$-$6128S&&&&&&&&&&\\
20551$-$4250&&&541.0&&582.2!&526.2!&481.2!&592.6!&&523.8&544.9&515.9&453.3\\
23128$-$5919N&&&629.8&&598.4!&&568!&578.4!&&466.0&336.3&451.1&318.7\\
23128$-$5919S&&&755.6&&760.3!&&719.4!&1832!&614.9&578.7&613.6&655.5&750.3:\\
23230$-$6926&&&&863.6&585.2&&639.9&714.6&1152.8&409.4:&578.2&606.3:&\\
23253$-$5415&&&&631.7&635.7&&&609.9&&&459.6&\\
23389$-$6139&&&803.9&1025.1&667.9&350.2&653.3:&817.0&486.2:&&674.5&336.4&672.2::\\
F23578$-$5307&&&&882.4&623.7&&&593.2&549.9&706.4:&713.9&\\
\hline
\end{tabular}
\\Col. (1) --- Name of Object.\\Col. (2) - Col. (14)
--- Observed full width at half maximum, in km/s. The instrumental line width
   is about 500 km/s.\\
! --- Line in region of poor atmospheric transmission between H and K band.

\end{table*}
\section{Results}
\label{Results}
The spectra for all 24 ULIRGs observed are shown in
Fig.~\ref{fig:spectra}.  The hydrogen recombination line $\Pa$ is
present in all ULIRG spectra except F00183$-$7111 where $\Pa$ is shifted
outside the K band. In most of the spectra, lines of the Brackett
series such as $\Brg$ at 2.166~$\mum$ and $\Brd$ at 1.945~$\mum$ are
observed as well.  Furthermore, in some ULIRGs we detected the helium
recombination line HeI (2$^{1}$S-2$^{1}$P$^{0}$) 2.058~$\mum$.  In
addition to the hydrogen recombination lines, most of the
ULIRG-spectra show strong rotation-vibration transitions of molecular
hydrogen, except F00183$-$7111 because of its redshift.  Transitions
between the vibrational first excited and ground state, 1-0 S(...),
dominate, suggesting that shocks or dense photo-dominated regions
(PDRs) are the main sources of molecular hydrogen emission
\citep{ste89}.  In some cases, lines of $\Fe$ with rest wavelengths
1.257 and 1.644~$\mu$m are observed in the H and/or K band.  In two
low redshift ULIRGs --- IRAS~20551$-$4250 and IRAS~23128$-$5919N/S --- CO
absorption bands at 2.29 and 2.32~$\mum$ are still present in the part
of the K band not yet strongly affected by the thermal background
radiation. These CO absorption bands are produced in the atmosphere of
red giants and supergiants.

\begin{figure*}[t]
\centering
\includegraphics[clip=true,scale=0.42]{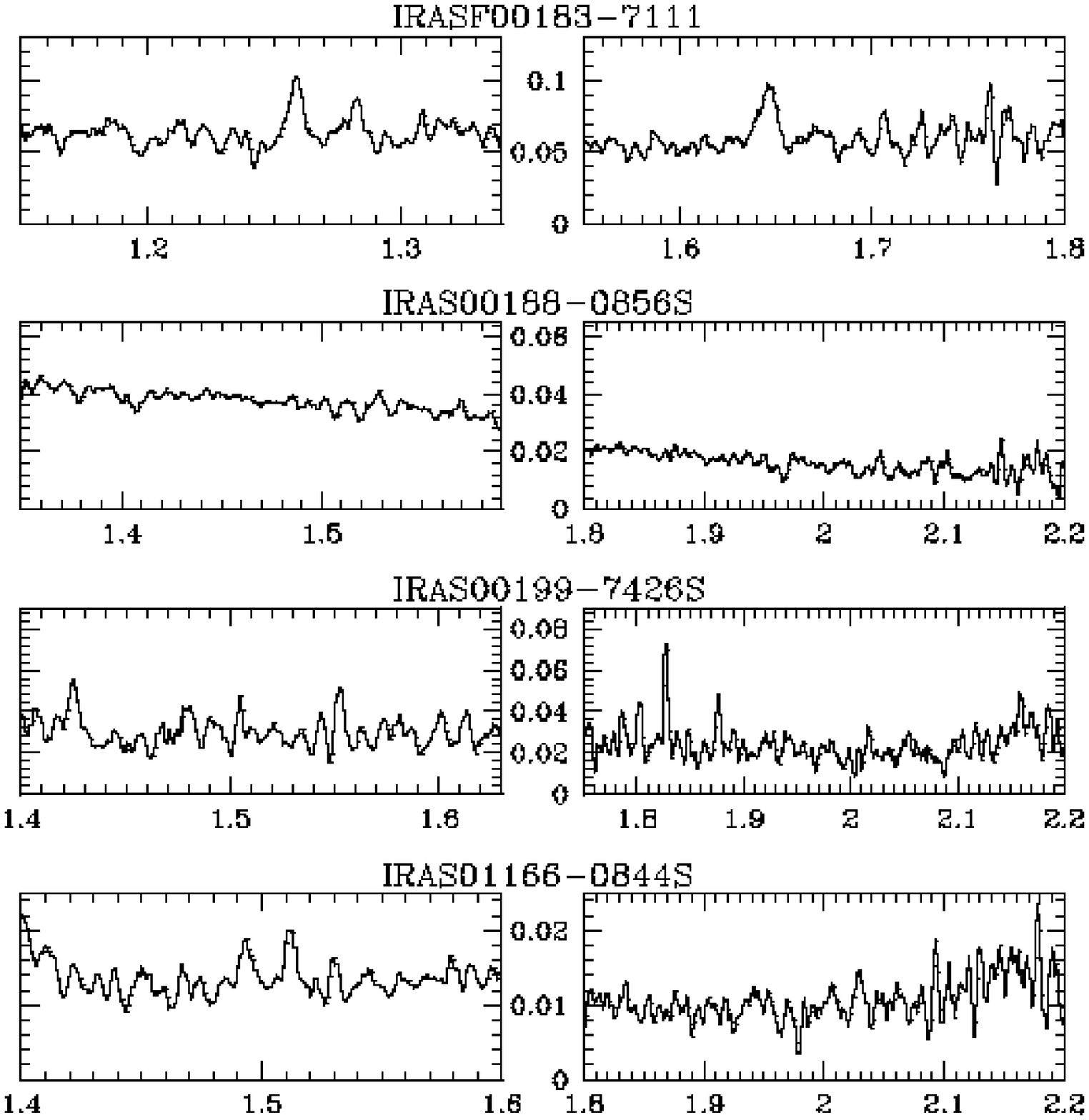}
\vspace{-.75cm}
\includegraphics[clip=true,scale=0.42]{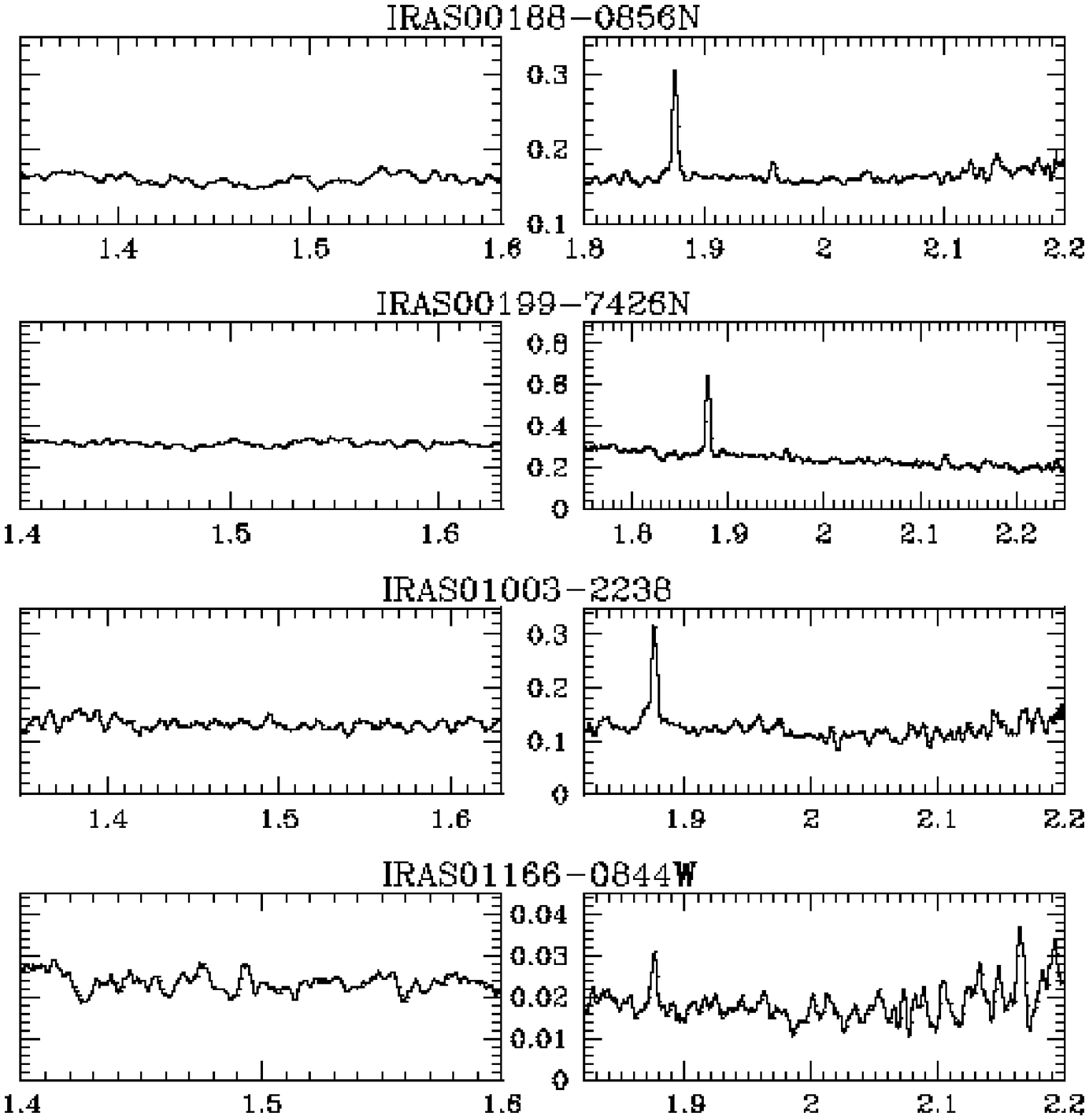}
\includegraphics[clip=true,scale=0.42]{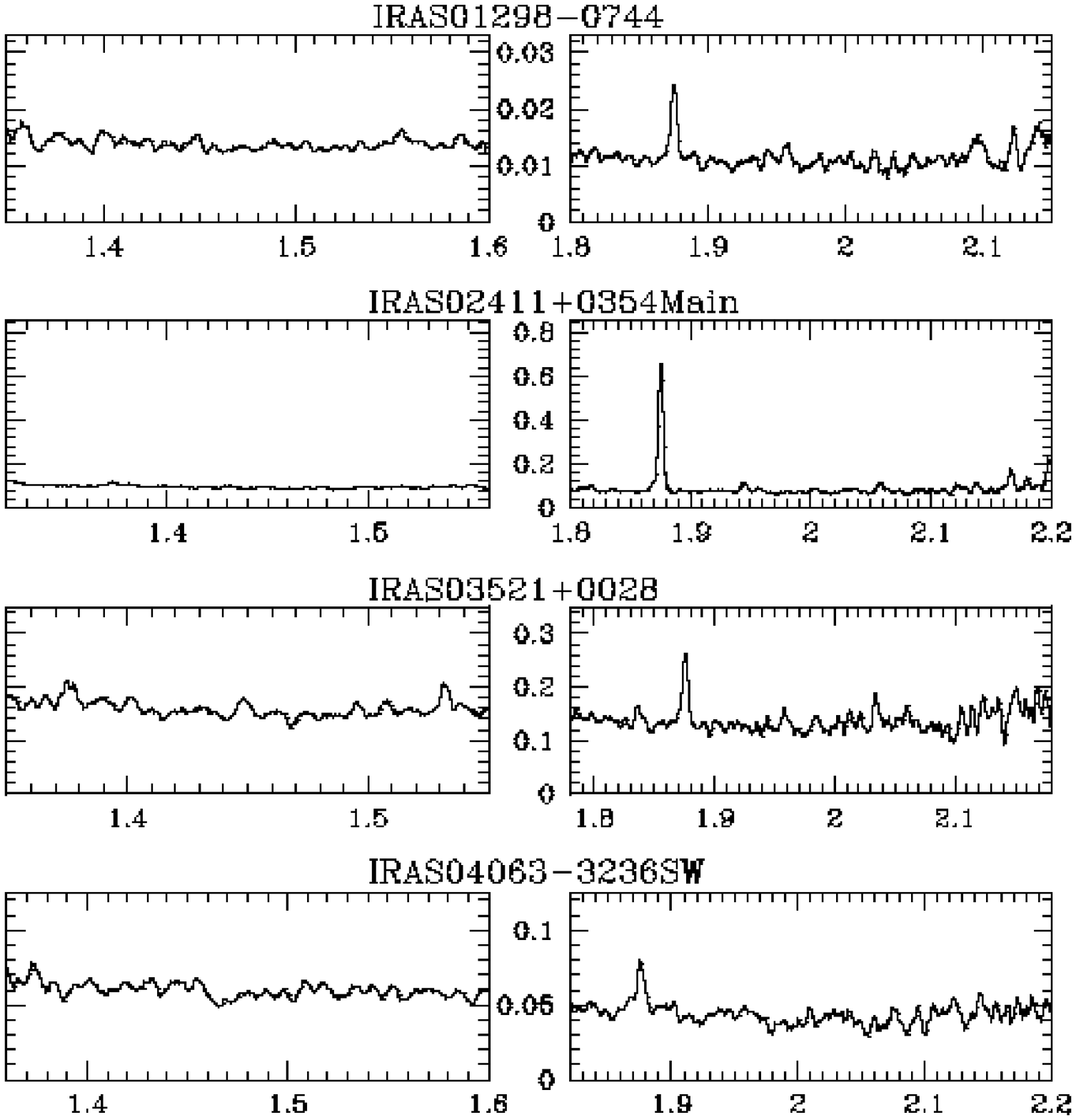}
\vspace{-.75cm}
\includegraphics[clip=true,scale=0.42]{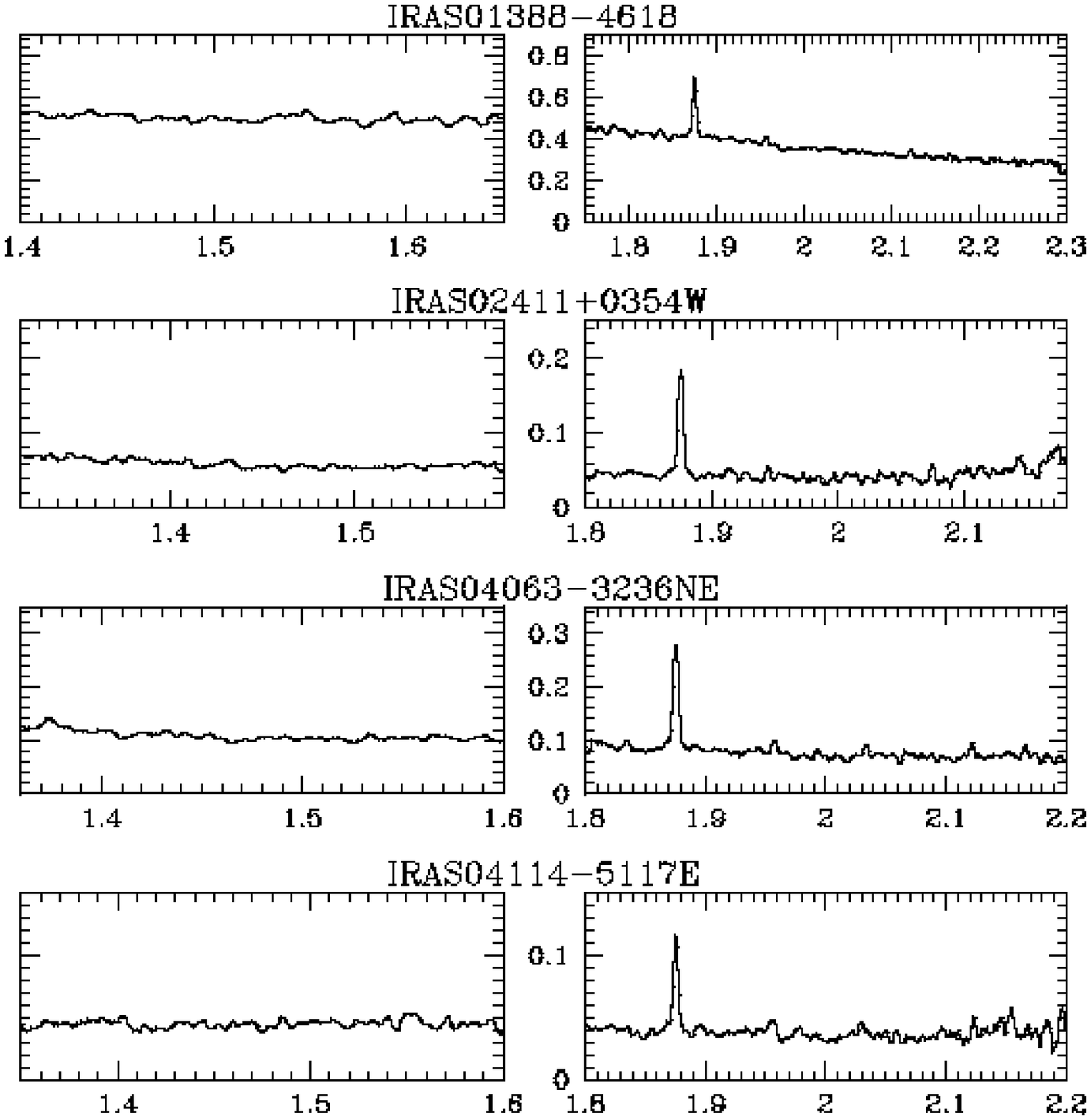}
\includegraphics[clip=true,scale=0.42]{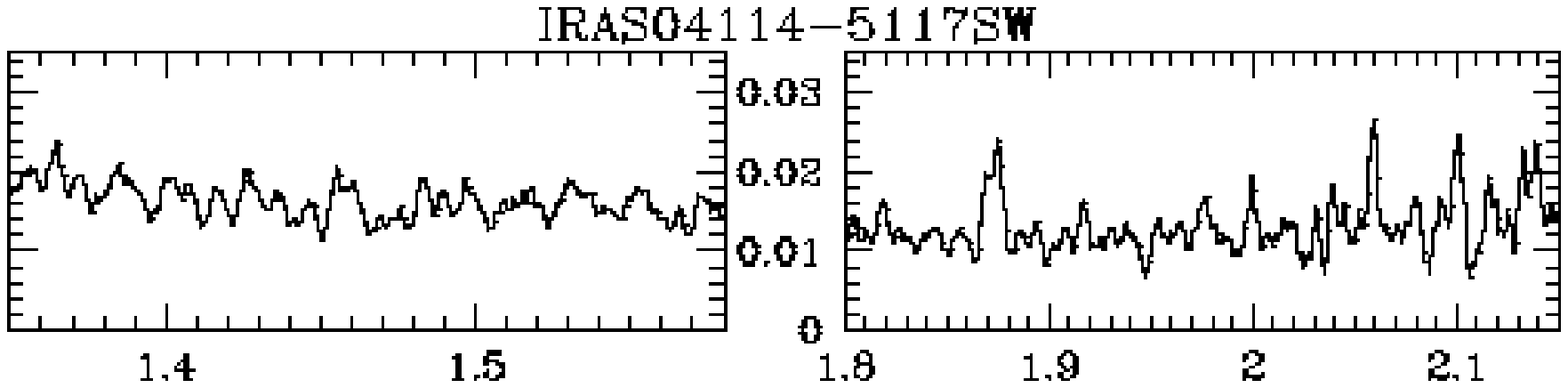}
\includegraphics[clip=true,scale=0.42]{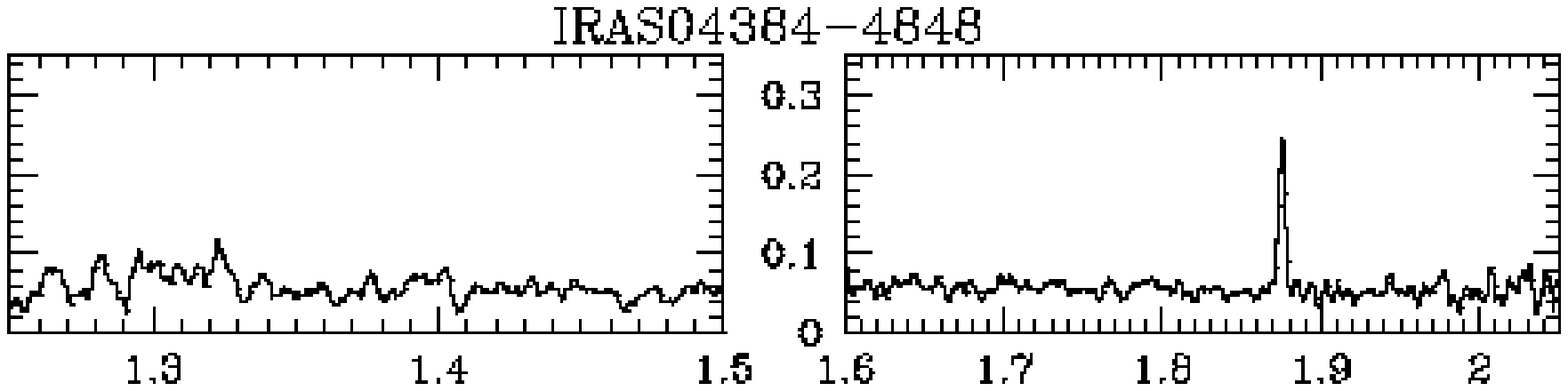}
\caption{ULIRG near-infrared spectra - flux density $f_{\lambda}$ as
function of the rest frame wavelength $\lambda$. For each object we
show in two panels the data taken in the H and K band and shifted to
the rest wavelength.  Flux density $f_{\lambda}$ is shown in units of
$10^{-11}$erg s$^{-1}$cm$^{-2}$$\mu$m$^{-1}$ and rest wavelength
$\lambda$ in $\mu$m.}
\label{fig:spectra}
\end{figure*}

\addtocounter{figure}{-1}
\begin{figure*}[t]
\centering
\includegraphics[clip=true,scale=0.42]{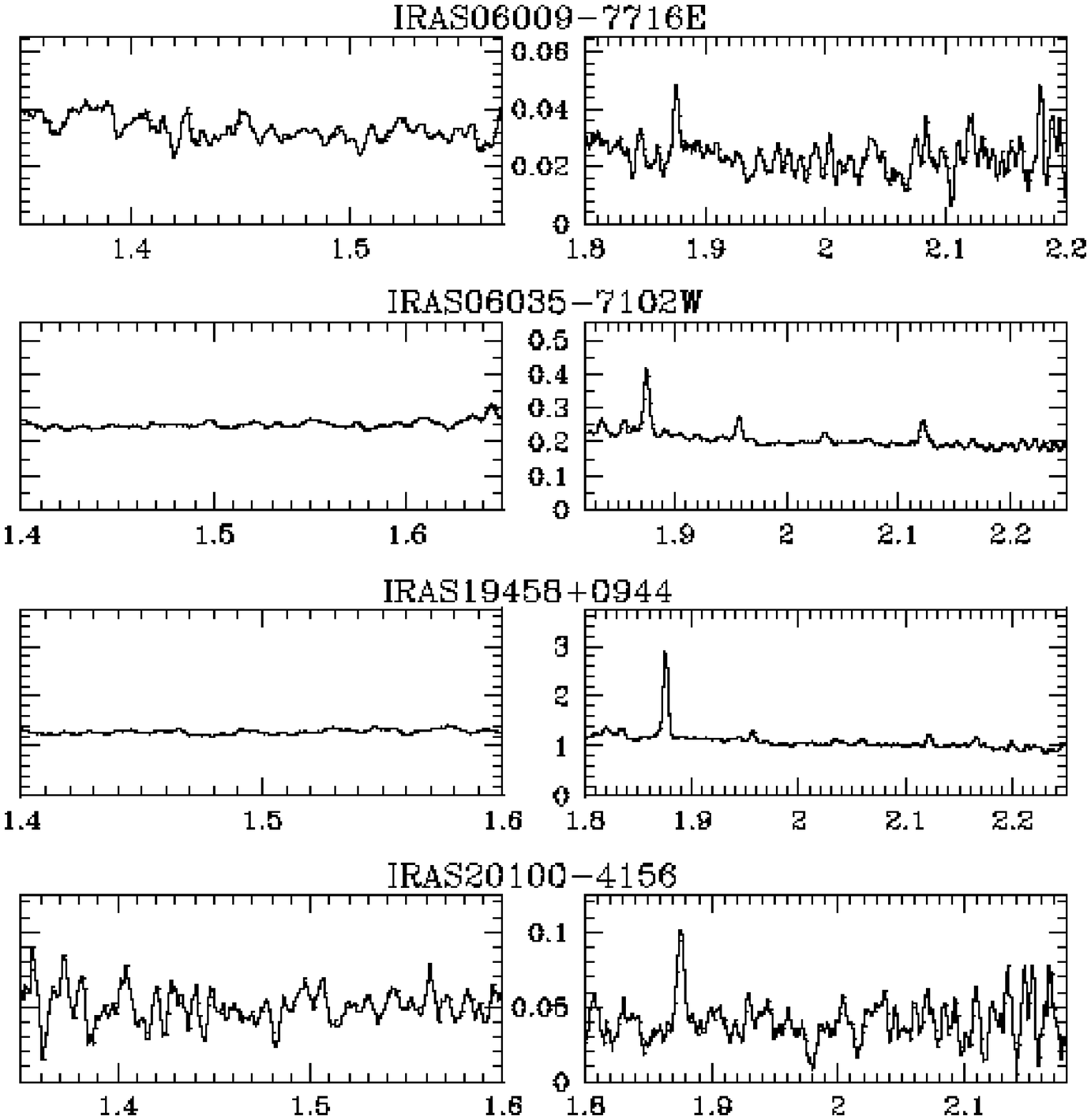}
\vspace{-.75cm}
\includegraphics[clip=true,scale=0.42]{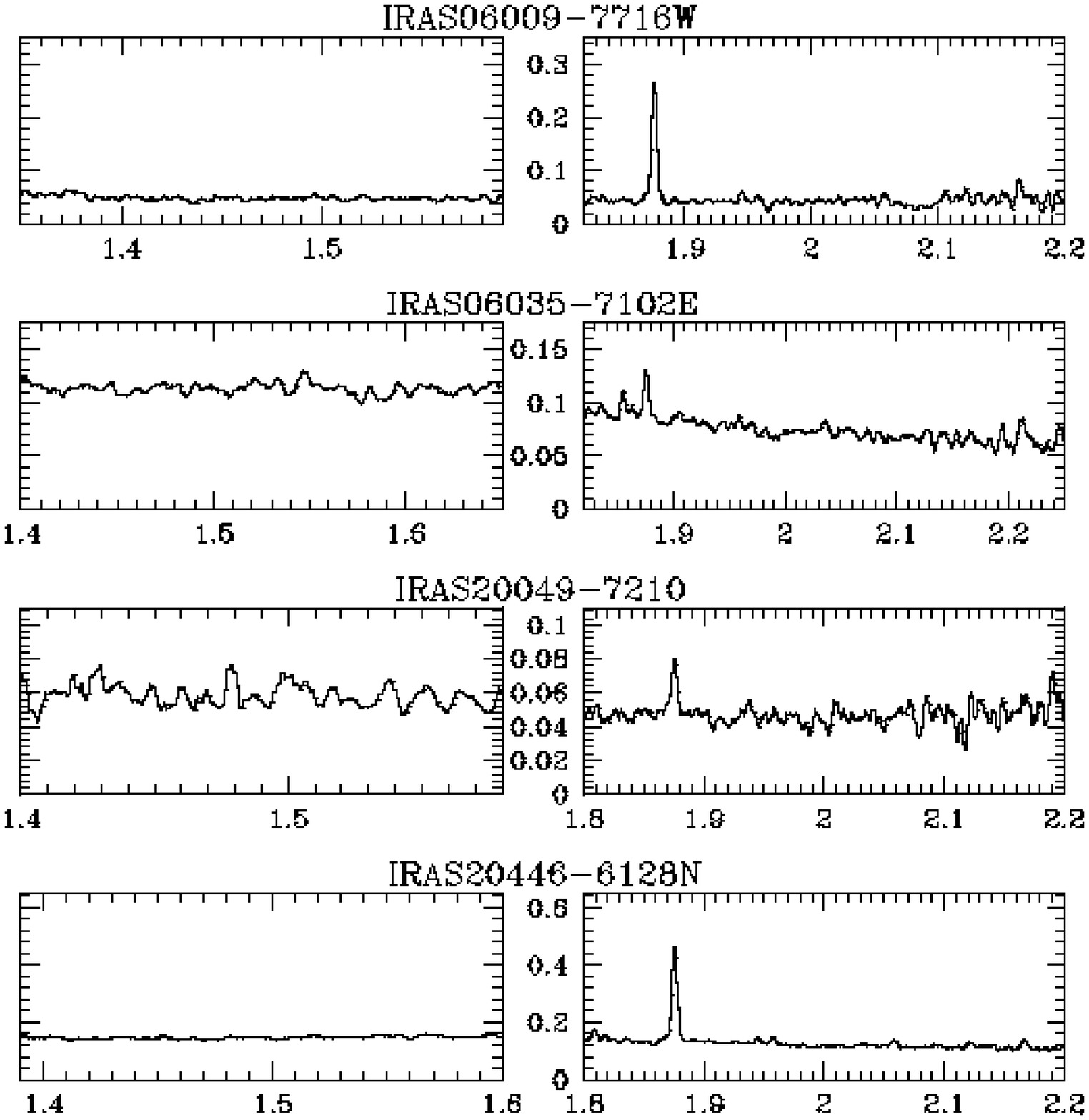}
\includegraphics[clip=true,scale=0.42]{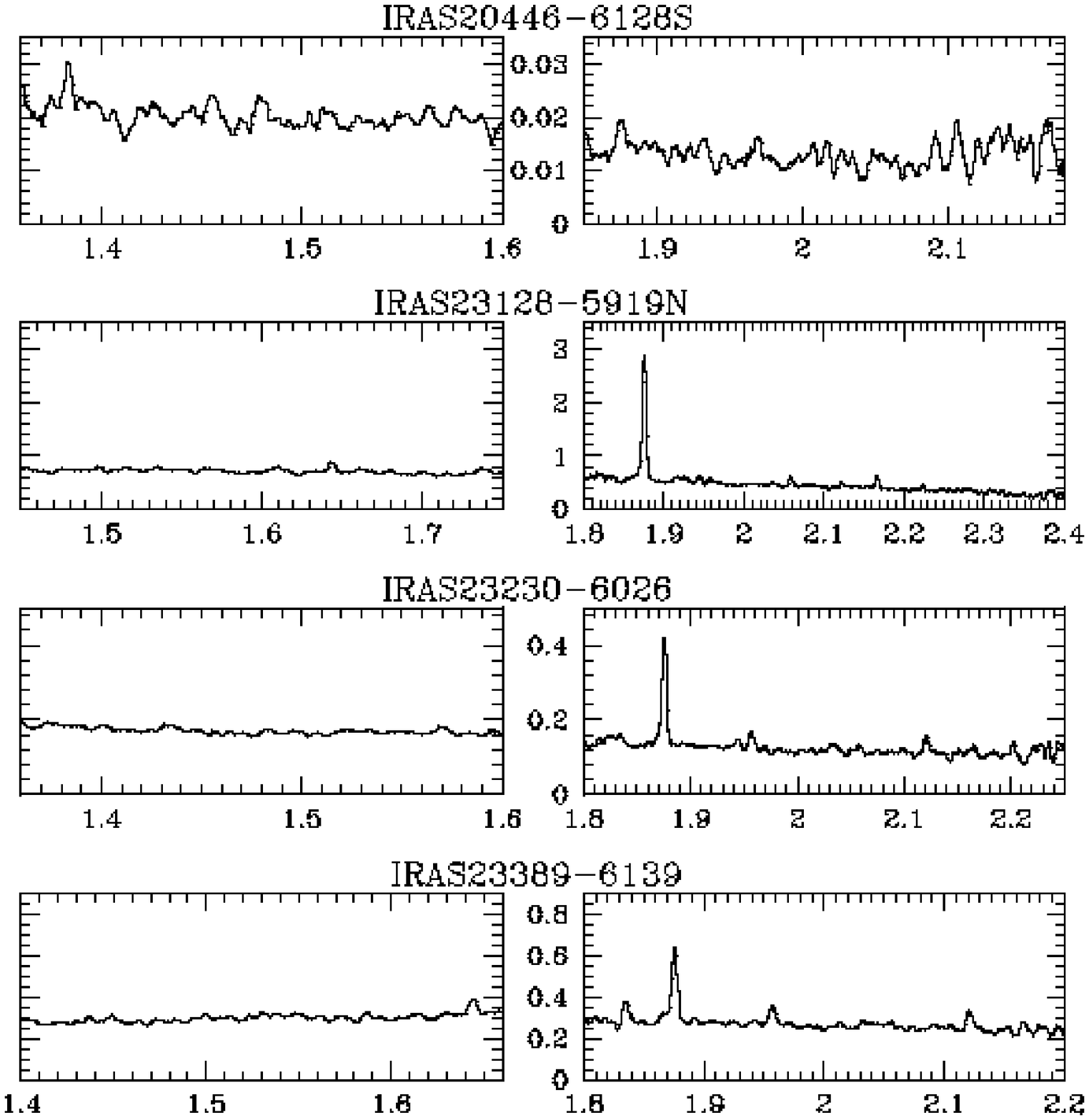}
\includegraphics[clip=true,scale=0.42]{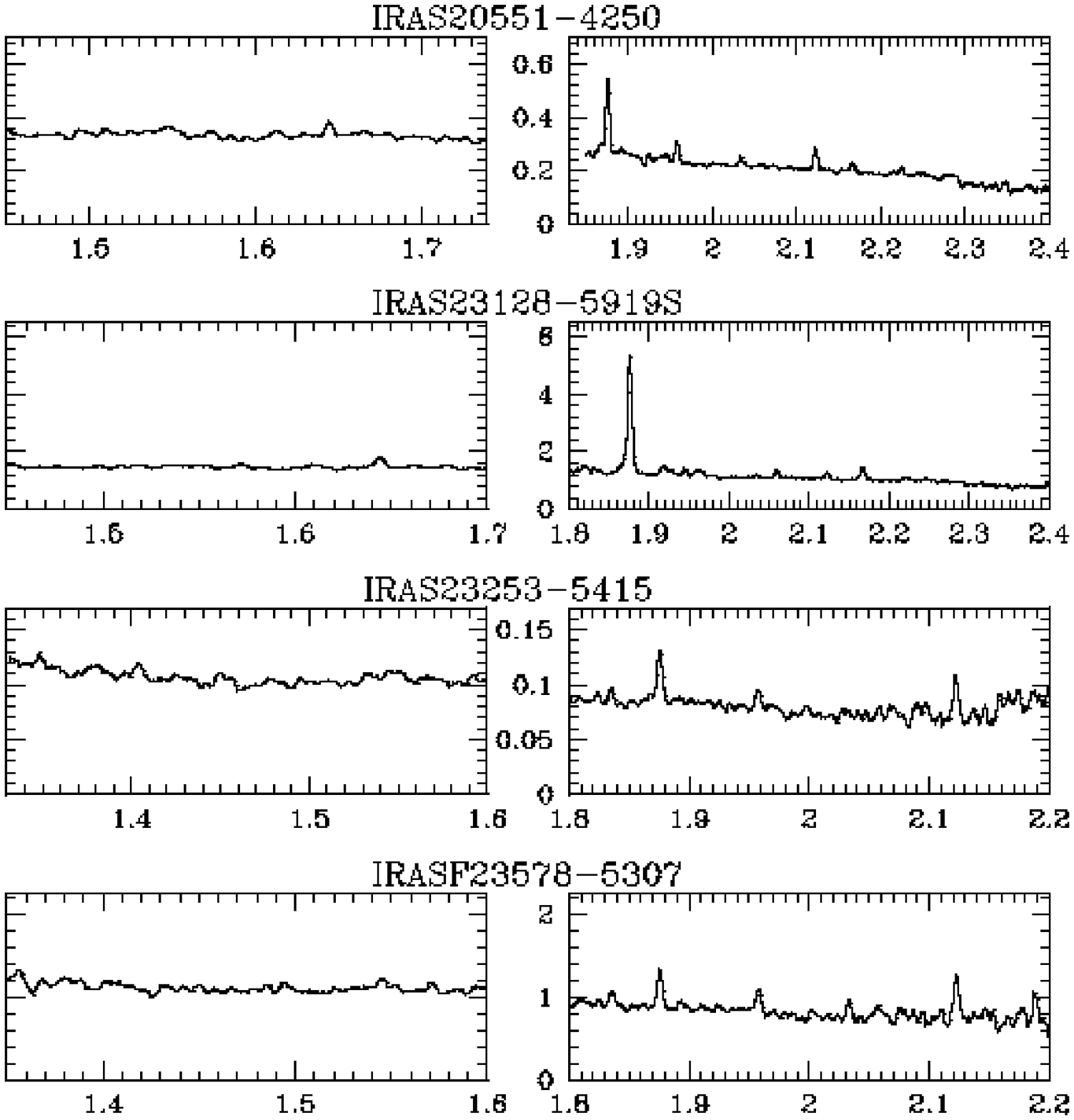}
\caption{(continued).}
\end{figure*}

\subsection{Search for buried AGNs}
\label{search}
The prime scientific motivation for our research is the search for `hidden
AGN' postulated by \citet{san88a}. Our aim is to compare near-infrared
evidence for AGN activity with evidence from ISO mid-infrared
observations of our sample ULIRGs \citep{rig99,tra01}.  Additionally,
we can quantify the extent to which near-infrared studies are affected
by extinction.  The near-infrared regime contains two indicators
linked to the presence of a `hidden AGN': $\Si$ and the hydrogen
recombination lines (e.g., $\Pa$, $\Pb$). \citet{mar94} showed that a
dust enshrouded AGN can be detected using the coronal $\Si$ line at
1.962~$\mu$m which is observed in most (though not all) local
Seyferts. The lower ionization potential of this species is 167 eV,
therefore, only the hard UV-spectrum of an AGN and not a starburst can
be responsible for the ionization up to Si$^{5+}$.  Spatially resolved
observations (e.g., \citealt{mai98}) and photoionization models of
e.g. ISO-observations \citep{ale99,ale00} are consistent with the
formation of this line in the inner part of the narrow-line
region. Near-infrared hydrogen recombination lines, especially $\Pa$
and $\Pb$, allow a deep search for AGN broad-line regions
(e.g., \citealt{vei97b}).  This method is obviously sensitive mainly to
type 1 AGN while, the BLR in type 2s remains usually undetectable out
to 4~$\mu$m \citep{lut02}. In contrast, [SiVI] emission from the
NLR can be detected for both type 1 and type 2 AGN.  Both methods
provide a further check of the mid-infrared method used by
\citet{rig99} and \citet{tra01} which is based on the ratio of large
scale PAH emission and AGN-heated dust.

Due to the SOFI spectral resolution of $\sim$ 600, $\Si$ is partially
blended with the molecular hydrogen line $\Hzweisdrei$ at
1.958~$\mum$. Models of fluorescent excitation of molecular hydrogen
predict an $\Hzweisdrei$/$\Hzweiseins$ flux ratio less than unity
(typically $\sim0.7$; \citealt{bla87}), while collisional excitation
models predict a ratio that ranges from 0.5 to 1.4
\citep{shu78,bla87,ste89}. Similar to \citet{vei97b,vei99b}, we have
assumed conservatively that $\Si$ is present in galaxies where the
apparent flux ratio of $\Hzweisdrei$/$\Hzweiseins$ is larger than 1.5.
IRAS~04114$-$5117 is a double nucleus system which most likely hosts an
AGN in one of its two nuclei.  In the East component
(IRAS~04114$-$5117E), the $\Hzweisdrei$/$\Hzweiseins$ ratio is 1.6,
consistent with the presence of an AGN. Also, the observed half width
of the blend of $\Si$ and $\Hzweisdrei$ lines in that source is higher
than the width of the $\Hzweiseins$ line alone (Table \ref{tab:hw}).
By fitting multiple Gaussians we have estimated the fluxes for the
various lines: values of $0.061\times10^{-14}$erg s$^{-1}$ cm$^{-2}$ and
$0.0165\times10^{-14}$erg s$^{-1}$ cm$^{-2}$ were found for the molecular
hydrogen and $\Si$ lines, respectively. The $\Si$ line flux renders an
observed luminosity of $1.4\times10^{6}$ L$_{\sun}$ for the $\Si$
line.  The presence of an AGN in this galaxy is also confirmed from
the mid-infrared classification using the PAH-diagnostic
\citep{rig99}.  The fainter component of this system,
IRAS~04114$-$5117W, shows no evidence of AGN activity.

For all sources, the hydrogen recombination lines $\Pa$ and $\Pb$, are
narrow with an intrinsic full width at half maximum (FWHM) of 700
$\kms$ at most (see Table~\ref{tab:hw} for the observed FWHM).  None
of the targets show evidence for an additional broad line component
($\Delta\upsilon$ a few 1000 km/s). In most of our ULIRGs, the $\Pa$
line shows a blue wing (see also
\citealt{vei97b,vei99b,bur01}). \citet{mur99,mur01} resolved similar
blue wings and proposed an identification with two helium
recombination lines at 1.8686 and 1.8697~$\mum$. Transitions of
molecular hydrogen may contribute as well \citep{vei99b}. Hence, we do
not consider such asymmetric blue wings as evidence for an AGN.

To better quantify the non-detection of a broad-line component in our
sample, we derived 3$\sigma$ upper limits for the flux of the
broad-line component of $\Pa$, assuming a line width of 3000
$\kms$. The line width of 3000 $\kms$ is a typical value for Seyfert 1
galaxies and consistent with the detections of broad $\Pa$ lines in
some ULIRGs by \citet{vei97b,vei99b}. From this, we determined an
upper limit for the ratio of the broad line flux and the measured
narrow-line flux.  A major part of our sample has a limit on this flux
ratio lower than one (21/29) and almost all (28/29) lower than
two. The non-detection of broad-line regions is significant
considering that ratios of broad and narrow flux significantly above 1 are
common in Sy1s \citep[e.g.,][where the median ratio is 20 for objects
not exceeding a luminosity of M$_{v}= -22$~mag]{sti90}, and have also
been observed in a few ULIRGs (IRAS~13451~$=2.3$, IRAS~23499~$=2.06$ and
IRAS~13305~$=0.86$, \citealt{vei97b,vei99b}).  IRAS~F00183$-$7111, which
was classified as AGN by ISO, cannot be classified unambiguously in the
near-infrared due to its redshift (precluding observation of $\Si$)
and faintness (precluding a good measurement or strong limit for a
broad component to $\Pb$).

Overall, we find evidence for AGN activity in one out of the 24 ULIRGs
of the present sample. Additionally, there is good agreement between
near-infrared and mid-infrared classifications.  None of the 17 ULIRGs
classified by ISO as starburst show any AGN activity in the
near-infrared. Small number statistics provided us with only 2 ISO AGN
in our sample --- one (IRAS~04114$-$5117E) also has near-infrared
evidence for AGN activity, the other (IRAS~F00183$-$7111) is uncertain.
This is consistent with the good agreement found between optical and
near-infrared \citep{vei97b,vei99b}, ISO and optical \citep{lut99} and
ISO and hard X-ray \citep{gen00} classifications, suggesting that
strong AGN in ULIRGs are often identifiable over a wide wavelength
range. Near-infrared and ISO diagnostics also agree in putting most
optical LINERs into the starburst category. This is true for three of the
four optical LINERs observed with SOFI, the fourth is again the
ambiguous IRAS~F00183$-$7111. As argued by \citet{lut99} and
\citet{tan99}, this suggests that optical LINERs from infrared
selected samples are linked to the starburst phenomenon.  We will
revisit this issue below when discussing $\Fe$ emission.

\subsection{Starburst luminosity and the `$\Brg$-Deficit'}
\label{Defizit}
Qualitatively, our results confirm previous findings that the majority
of near-infrared spectra of ULIRGs are
starburst-like. \citet{gol95,gol97a,gol97b} studied quantitatively in
the near-infrared whether the starburst activity traced is sufficient
to power the bolometric luminosity of infrared galaxies. To quantify
the relative strength of the ionization source, they used the
$\Brg$-luminosity (which is proportional to the ionizing photon rate)
in conjunction with the infrared luminosity $\Lir$ which is assumed to
be equal to the bolometric luminosity of the (U)LIRGs.  $\Brg$ has
been extinction corrected using either near-infrared continuum colors
or E(B$-$V) derived from the optical hydrogen recombination lines
$\Ha$ and $\Hb$. Goldader et al., found that in luminous infrared
galaxies (LIRGs), the starburst activity seen in the near-infrared can
account for the total bolometric luminosity. In ULIRGs, however, they
found that the starburst activity detected is insufficient to account
for the bolometric luminosity if similar starburst properties (initial
mass function, star formation history) to those for the LIRGs are
assumed.  This `deficit' in $\Brg$ luminosity or ionizing luminosity,
is usually a factor of about 3, although it can reach up to 10 in
individual cases. A likely interpretation is that a significant part
of the ULIRG's power source has not yet been seen in these
observations because the central regions are optically thick in the
near-infrared.  \citet{vei97b,vei99b} have investigated the deficit of
the ionizing luminosity using the $\Pa$ line. They determined
extinction (in a foreground screen model) using optical/optical and
optical/near-infrared hydrogen recombination-line pairs. They confirm
the deficit found by \citet{gol95,gol97a,gol97b} for ULIRGs that are
cool in the IRAS 25~$\mum$/60~$\mum$ color but do not find a deficit
for warm\footnote{
\cite[f$_{\nu}$(25~$\mu$m)/f$_{\nu}$(60~$\mu$m)$>$0.2;][]{san88b}}
(mostly AGN-like) ULIRGs, a category excluded by
\citet{gol95,gol97a,gol97b} in their initial analysis. Recent work by
\citet{val05} report this deficit using the $\Pa$ and $\Brg$ line as well.

We used the dominant $\Pa$ line to revisit the deficit of the ionizing
luminosity for our sample of mostly starburst-like and `cool' ULIRGs.
To determine the intrinsic flux of this line, we corrected for
extinction using the foreground absorber model. We adopted the
extinction coeffcients of \citet{vei97a} and three different line
pairs: $\Ha$/$\Hb$, $\Pa$/$\Ha$ and $\Pa$/$\Brg$.  For the
calculations we assumed case B recombination, a temperature $T \sim
10^{4}$ K and an electron density $n_{e} \sim 10^{4}cm^{-3}$
\citep{hum87}. Different line pairs often provide significantly
different extinction values.  In principle, the extinction derived
from the purely near-infrared line pair $\Pa$/$\Brg$ is significantly
higher than that from the purely optical $\Ha$/$\Hb$ or the
optical/near-infrared $\Pa$/$\Ha$, indicating that the near-infrared
recombination-line emission originates in deeper more obscured regions
and does not follow a simple screen model. Technical reasons like
different aperture sizes may also affect the line ratios derived from
independent optical and near-infrared observations. On the other hand,
$\Pa$ and $\Brg$ are separated by less than 0.3~$\mum$, and thus give
useful extinction constraints only for high A$_V$ ($\geq$10) because
of the small differential extinction. For some of our sources, $\Brg$
is redshifted into the less favourable long end of the K band. Since
we are probing for obscuration effects that are significant even in
the near-infrared, and since both lines are taken from the same
spectrum, thus minimizing systematic errors, the $\Pa$/$\Brg$ ratio is
valuable for studying the ionizing photon deficit in ULIRGs.

\begin{figure}
\resizebox{\hsize}{!}{\includegraphics{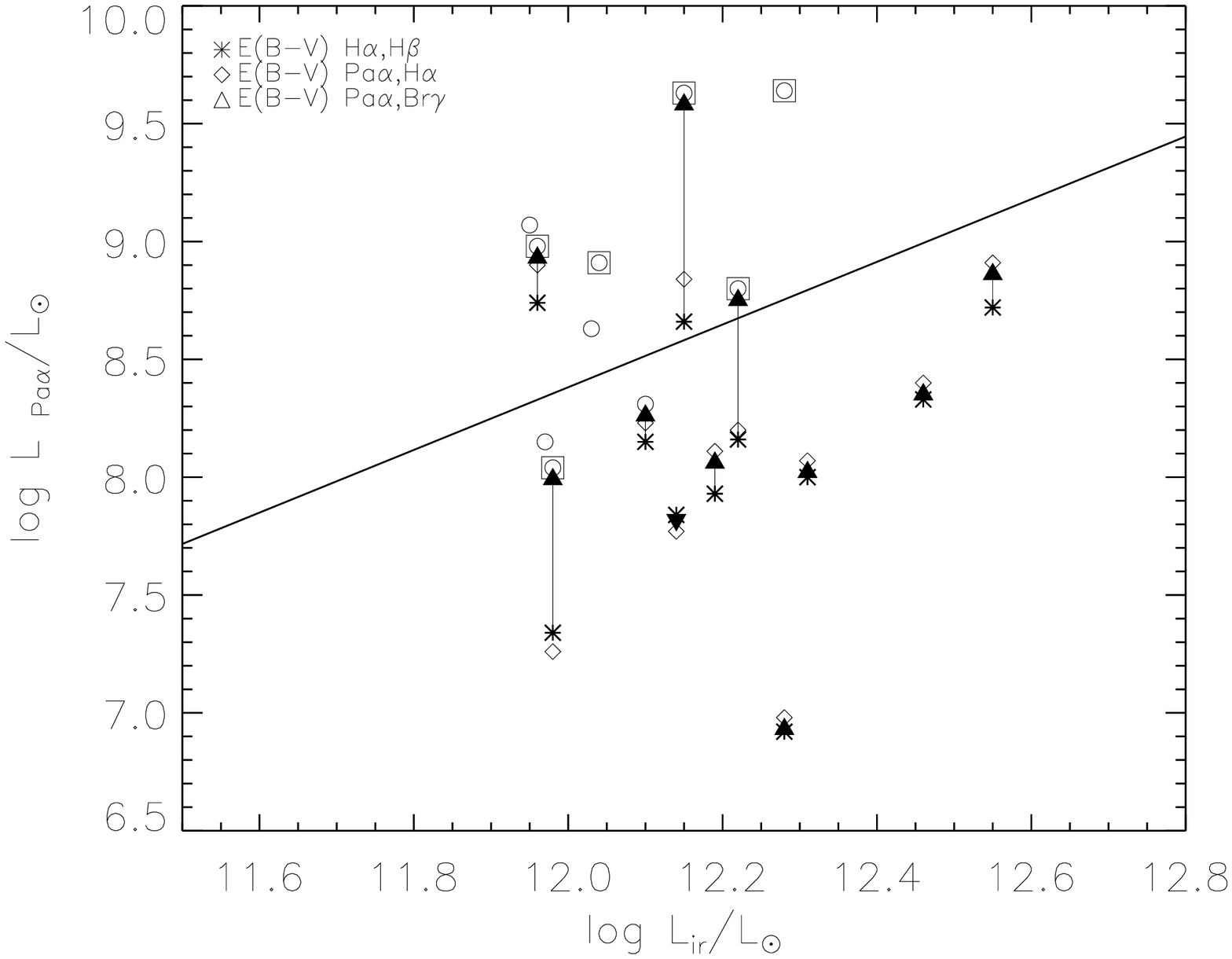}}
\caption{\label{fig:deficit}Logarithm of the Pa$\al$-luminosity plotted as a function of
the logarithm of the far-infrared luminosity $\Lir$. The solid line
represents the relation derived by \citet{gol95,gol97a,gol97b} for
starburst LIRGs, transformed to $\Pa$ assuming case B
ratios. The different symbols represent the different
extinction
corrections used to calculate the Pa$\al$-luminosity $\LPa$. The arrows
visualize the change for extinction corrections done at progressively longer
wavelengths. ULIRGs in which HeI 2.058~$\mum$ is detected are marked by a square.}
\end{figure}

In Table~\ref{tab:extinction} we summarize the results for those
sample galaxies for which we estimated extinctions, and give
references to the optical spectra used.  In Figure~\ref{fig:deficit}
we plot the $\Pa$ luminosities (corrected for extinction based on
estimates from the various line pairs) as a function of infrared
luminosity.

\begin{table}
\caption{Colour excess E(B-V) as derived from different
recombination line pairs\label{tab:extinction}}
\begin{tabular}{lccccc}
\hline\hline
Object&E(B-V)&E(B-V)&E(B-V)&Ref.\\
&$\frac{\Ha}{\Hb}$&$\frac{\Pa}{\Ha}$&$\frac{\Pa}{\Brg}$&\\
(1)&(2)&(3)&(4)&(5)\\
\hline
00188-0856N&1.46&1.59&&1\\
01003-2238&0.71&1.43&&1\\
01166-0844&0.32&&&1\\
01298-0744&0.71&0.77&&1\\
01388-4618&&&4.64\\
02411+0354Main&0.66&1.41&6.37&1\\
02411+0354W&0.79&1.07&&1\\
03521+0028&2.36&2.43&&1\\
04063-3236NE&&&1.85\\
04063-3236SW\\
06009-7716E&&&\\
06009-7716W&&&6.40\\
06035-7102W&1.36&0.85&&2\\
06035-7102E&0.94&0.08&&2\\
19458+0944&&&5.54\\
20100-4156&0.96&1.75&\\
20446-6128N&&&5.23\\
20446-6128S\\
20551-4250&1.06&0.48&4.88&2\\
23128-5919N&1.03&1.45&1.19&2\\
23128-5919S&0.76&1.77&2.46&2\\
23230-6926&1.30&1.28&4.71&2\\
23389-6139&1.01&1.19&1.67&1/2\\
\hline
\end{tabular}
\\ 
Col. (5) --- References for the optical line measurements. 1: \citet{vei99a}. 2:
\citet{duc97}.\\
The calculations are based on case B recombination, a temperature
$T\sim10^{4}$ K, an electron density $n_{e}\sim10^{4}cm^{-3}$ \citep{hum87} and the foreground
absorber modell. We adopted the extinction coefficients of \citet{vei97a}.
\end{table}

Our main results are: For most of the ULIRGs with extinction
correction based on purely optical $\Ha$/$\Hb$ and
optical/near-infrared $\Pa$/$\Ha$, we detect a deficit of $\Pa$ flux
with respect to what one would expect for a starburst (solid line in
Fig.~\ref{fig:deficit}), thus confirming the results of
\citet{gol95,gol97a,gol97b}. 
\newpage
For the subsample where the extinction
correction came from purely near-infrared $\Pa$/$\Brg$ line ratio,
there is definitely no $\Pa$-deficit but rather on average an excess
of $\Pa$ flux. It is difficult to estimate to which extent this excess
would be reduced if a similar near-infrared extinction correction were
to be applied to the lower luminosity reference sample --- the
reference objects are typically dusty starbursts for which such
`wavelength-dependent' extinction may be present as well, at a lower
level than in the ULIRGs (see the case of M~82 \citep{mcleod93} which
also shows that this `wavelength-dependent' extinction is an artifact
of enforcing a screen model).  A tendency may be present in our data
for galaxies with high $\Pa$/IR to have detections of the HeI line at
2.058~$\mum$ (Fig.~\ref{fig:deficit}). Despite various other complex
effects in the interpretation of this line (e.g., \citealt{shields93}),
this may be consistent with younger bursts having both harder ionizing
radiation and a higher ratio of ionizing to bolometric luminosity. In
agreement with the results from ISO spectroscopy \citep{gen98}, our
findings suggest that the deficit in ionizing luminosity derived from
near-infrared spectra of ULIRGs is indeed an obscuration effect, such
that much of the `missing' luminosity is in the form of highly
obscured (A$_V\approx10-20$ mag) starburst activity.

\subsection{$\Fe$ Emission in ULIRGs: Characteristic of the LINER-phenomenon?}
\label{Emission}
\begin{figure}
\resizebox{\hsize}{!}{\includegraphics{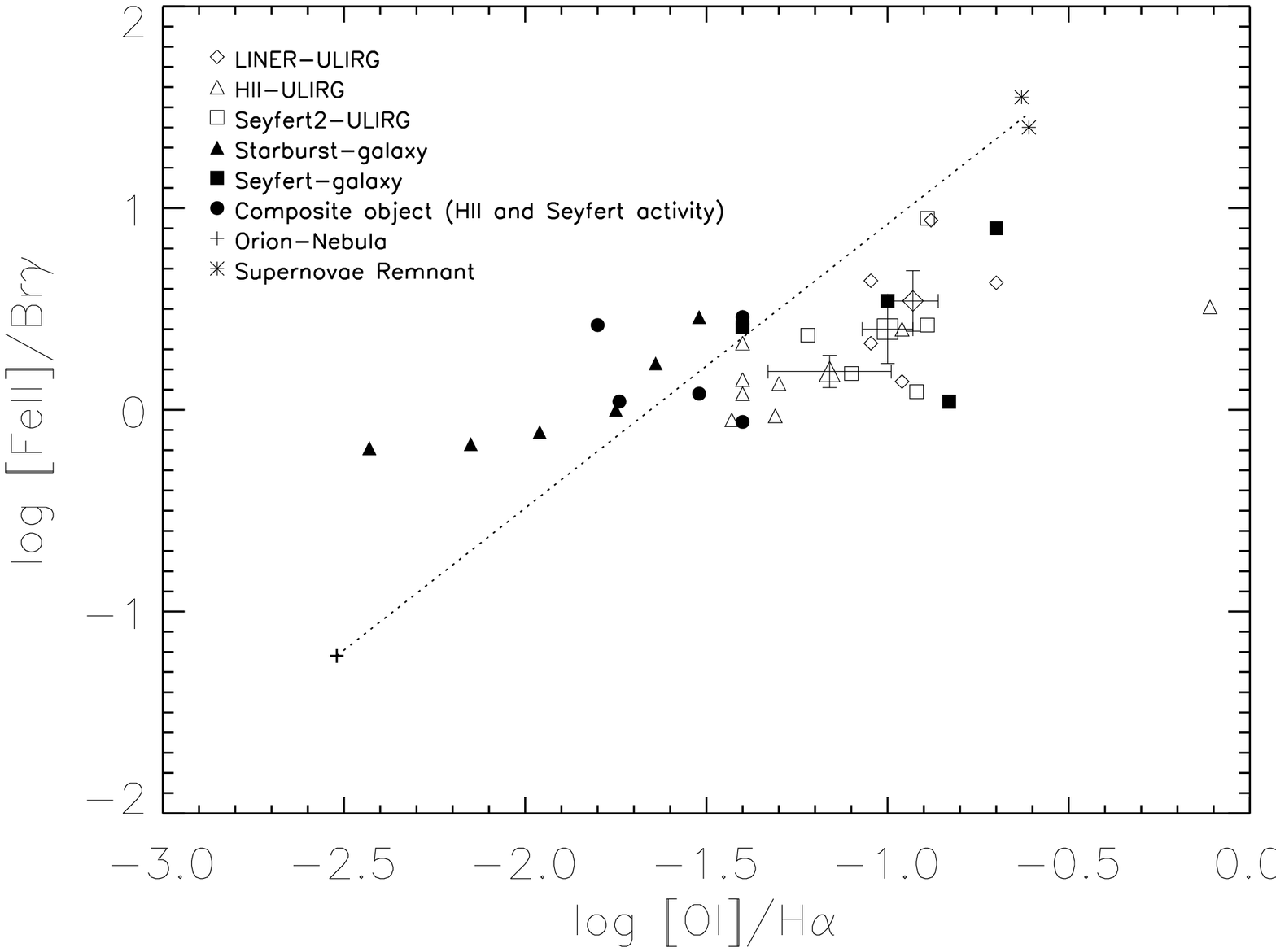}}
\caption{\label{fig:fe}Logarithm of the flux ratio ${\Fsechs}/{\Brg}$ versus
the logarithm of the flux ratio ${\Oi}/{\Ha}$ both for ULIRGs and
lower luminous emission-line galaxies. The enhanced symbols with error
bars represents the average of each ULIRG sub-sample.
The dotted line
shows a schematic mix of the Orion-nebula (HII-region) and the
supernova remnants IC~443 and  RCW~103 (shocked sources).}
\end{figure}

\citet{rie80} detected for the first time a near-infrared line of
$\Fe$ at 1.644~$\mum$ ( $a^{4}F_{9/2}-a^{4}D_{7/2}$) in an
extragalactic object, the starburst galaxy M~82. A few years later, by
investigating the supernova remnant IC~443, \citet{gra87} proposed that
this line could be a good indicator for shock activity in astronomical
objects. \citet{mou90} and \citet{mou93} pointed out a correlation
between $\Fsechs$ and $\Oi$ ascribed to the fact that $\Fsechs$ and
$\Oi$ arise mostly in partially ionized zones which are created either
by relatively fast shocks (100 $\kms$) or by X-ray radiation from an
AGN. They established a dignostic diagram, $\Fsechs$ and $\Oi$ being
normalized by the hydrogen recombination lines $\Brg$ and $\Ha$,
respectively (Fig. 1a in \citet{mou90} and Fig.~\ref{fig:fe} in this
paper). As the hydrogen recombination lines arise from fully ionized
zones, the ${\Fsechs}/{\Brg}$ and ${\Oi}/{\Ha}$ ratios are a
measure of the ratio of the volume of the partially relative to the
fully ionized zones.  Figure~\ref{fig:fe} shows the two line ratios:
such a diagram sketches nicely the increasing importance of shocks, or
of X-ray excitation when going from star-formation to AGN activity.
Several studies (e.g. \citealt{mor88,gre91,gre97,mou93,for93,van97})
argued that in the case of starburst galaxies shock activity caused by
supernova remnants and superwinds is responsible for the observed
$\Fe$ emission and $\OI$ emission.

Our SOFI spectra cover most of the H and K band, including [FeII] at
1.644 and/or 1.257~$\mum$ for several targets. This motivated a study
of the $\Fe$ properties of ULIRGs in general for which we added ULIRGs
from \citet{vei97b,vei99b} in order to improve the
statistics. Our sample includes ULIRGs with both $\Oi$ and $\Ha$
measurements \citep{vei95,vei97b,vei99b,duc97,kim98}. In total our
sample comprises 21 ULIRGs.  We categorize the ULIRGs according to
their optically classified spectral type as Starburst-, AGN- or
LINER-ULIRGs following the classification scheme established by
\cite{vei87}.  In addition, we include in Fig.~\ref{fig:fe} for
comparison, some lower luminosity Starbursts, Seyfert and composite
galaxies. Representatives of pure shock- and photoionization
excitation, the supernova remnant IC~443 and RCW~103 as well the Orion
nebulae are added
\footnote{Data of reference objects are taken from:
\citet{mor88}; \citet{arm89};
\citet{ver86}; \citet{kaw88}; \citet{ait81}; \citet{fre80}; \citet{joh87};
\citet{rie80}; \citet{pei70}; \citet{rie81}; \citet{bok75}; \citet{wel70};
\citet{wil76}; \citet{les88}; \citet{hec87}; \citet{coh83}; \citet{low79};
\citet{bal80}; \citet{gra87}; \citet{fes80}; \citet{oli89}; \citet{lei83};
\citet{vei97b,vei99b}; \citet{vei99a}; \citet{duc97}; \citet{kim98}}.

Since integrated ULIRG spectra represent a heterogeneous mixture of
different astronomical objects, (e.g., HII-regions, SNRs, PDRs, winds
and AGNs) it is not surprising to find them between the loci of pure
HII-regions and pure shock excitation. Starburst-ULIRGs and
LINER-ULIRGs populate different regions of the diagram. On average
(see Table~\ref{tab:fe-average} and in Fig.~\ref{fig:fe} the enhanced
symbols with error bars), both flux ratios, ${\Fsechs}/{\Brg}$ and
${\Oi}/{\Ha}$, are higher in LINER-ULIRGs than in
starburst-ULIRGs. The latter also have somewhat higher ratios when
compared to lower luminosity starbursts. AGN-ULIRGs populate a region
of high ratios, similar to lower luminosity Seyferts.  The complex
origin of the iron emission makes clear statements difficult, in
particular for individual sources.
\begin{table}
\caption{Mean flux ratios ${\Oi}/{\Ha}$ and ${\Fsechs}/{\Brg}$
for subsamples of ULIRGs and emission-line galaxies\label{tab:fe-average}}
\begin{tabular}{lcc}
\hline\hline
Object class&log\{${\Oi}/{\Ha}$\}&log\{${\Fsechs}/{\Brg}$\}\\
(1)&(2)&(3)\\
\hline
LINER-ULIRG&-0.93$\pm$0.07&0.54$\pm$0.15\\
HII-ULIRG&-1.16$\pm$0.17&0.19$\pm$0.08\\
Seyfert2-ULIRG&-1.00$\pm$0.07&0.40$\pm$0.17\\
HII-galaxy&-1.85$\pm$0.15&0.10$\pm$0.10\\
Seyfert-galaxy&-0.98$\pm$0.18&0.47$\pm$0.21\\
Composite object$^{\ast}$&-1.57$\pm$0.09&0.19$\pm$0.12\\
\hline
\end{tabular}
\\$^{\ast}$: Composite objects exhibit both starburst and
  seyfert activity.
\end{table}

In what follows we describe a scenario which could be a plausible
explanation of our observations: All starburst-ULIRGs and even more
strongly the LINER-ULIRGs exhibit an elevated level of supernova- or
superwind-related shock activity when compared to normal (lower
luminosity) starbursts.  A number of effects may contribute to this
behaviour: \citet{alo00}, among others, suggested that the elevated
shock activity (with respect to HII-region activity) could be due to
an aging starburst. The hot ionizing stars live only for about 10
million years while supernova and wind activity sets in after a few
million years and persists longer. The detection of the HeI
2.058~$\mum$ line in some starburst-ULIRGs but no LINER-ULIRGs is
indeed consistent with this model. Furthermore, the measured
equivalent width of the hydrogen recombination lines being an
indicator of the age of the burst tends in LINER-ULIRGs to be smaller
than in HII-ULIRGs. Applying on our sample Fig.~7 of \citet{alo00}
where log\{${\Fsechs}/{\Brg}$\}\footnote{Following \citet{sim96}, this
flux ratio could be used as a rough estimator for the age of the
starburst.} versus log\{K/$\Brg$\} is plotted, we would estimate burst
ages between $8-12$~Myrs for the LINER-ULIRGs (assuming a burst
duration of 1~Myr).  Additionally, differential extinction could also
play a role: a larger scale wind will be less obscured than the
embedded star forming regions. \citet{gol95} invoked this scenario to
explain the {\em absence} of a deficit in H$_2$ emission in ULIRGs
showing a $\Brg$ deficit. Qualitatively, such an effect may be
indicated by sources having shock-like ${\Oi}/{\Ha}$ ratios (from the
outer wind) but more HII-region like ${\Fsechs}/{\Brg}$ (from the more
obscured starburst).  Figure~\ref{fig:fe} may indeed contain such
sources.

AGN-ULIRGs populate a similar region in Fig.~\ref{fig:fe} as do lower
luminosity Seyferts. This likely means a strong contribution of AGN
X-rays and AGN winds to their $\Fe$ emission, but the considerable
overlap with other categories also allows significant non-AGN
contributions.  It is unlikely that AGN dominate the $\Fe$ emission of
the starburst- and LINER-ULIRGs. It would be difficult to consistently
avoid detection of $\Si$ from the AGN in that case. Most of the
observed $\Fe$/$\Si$ limits of the starburst-ULIRGs are higher than
the ratios for bona fide Seyferts. This indicates that [FeII] emission
in starbursts-ULIRGs and LINER-ULIRGs likely originates in a starburst
superwind rather than in an AGN.

\subsection{Notes on individual galaxies}
\label{extraordinary objects}
\subsubsection{F00183$-$7111: Extreme $\Fe$ Emission}
\label{F00183}
F00183$-$7111 is the most distant (z~$=0.32$) and most luminous object
($\Lir = 10^{12.77}$ L$_{\sun}$) in the ULIRG-sample observed with
SOFI. \citet{arm89} classified this galaxy in the optical as a
LINER. In the mid-infrared, \citet{tra01} found that this object
displays extraordinary absorption features of silicates and ices and
interpreted it as a heavily obscured AGN. Recently, Spitzer IRS
observations by \citet{spo04} strengthened this interpretation. Unfortunately, the
prime AGN indicator $\Si$ is redshifted outside of the K band and
cannot be used. We do not see a broad $\Pb$ component. In this ULIRG,
we measured the greatest intrinsic line width of [FeII] lines in our
sample, $\sim900\kms$ and $\sim1000\kms$, for $\Fzwei$ and $\Fsechs$,
respectively.  The flux ratio of ${\Fsechs}/{\Brg}=0.94$ (from
measured $\Pb$) is higher than in normal ULIRGs and similar to the
'near-ULIRG' NGC~6240. Both an AGN and a shock origin of this emission
are possible, the large intrinsic half width of the iron lines, larger
than $\Pb$, may favour shock activity.  The observed luminosity of
both lines combined is $3.7 \times 10^{8}$ L$_{\sun}$. For
comparision, in NGC~6240 the observed luminosity of these two lines
combined is $4.9\times 10^{7}$ L$_{\sun}$.  The extinction to the
$\Fe$ emitting region can be estimated using the two iron lines
arising from the same upper level, with an intrinsic flux ratio
${\Fzwei}/{\Fsechs}$ of 1.36 \citep{nus88}.  We obtain
$A_{H}=1.45$ mag using the extinction curve of \citet{rie85}, and
$A_{H}=1.2$ mag using \citet{kor83}.  Assuming a foreground absorber
we estimate an extinction corrected luminosity of both lines combined
of $2 \times 10^{9}$ L$_{\sun}$ and $1.5\times 10^{9}$ L$_{\sun}$ for
the two extinction laws, respectively. The extreme $\Fe$ luminosity
and the unusual mid-infrared spectrum \citep{tra01,spo04} motivate further
studies of F00183$-$7111 as an atypical ULIRG.

\subsubsection{F23578-5307: The galaxy with one of the highest $\Hzweiseins$ luminosities}
\label{F23578}
The $\Hzwei$ lines of the rotation-vibration transition 1-0 S(...)
dominate the spectrum of F23578$-$5307 (z~$=0.125$; $\Lir = 1.3\times
10^{12}$ L$_{\sun}$) and, atypical for ULIRGs, $\Pa$ is not the
strongest line in the near-infrared spectrum. Very few ULIRGs have
similar characteristics, the best known example being the `near-ULIRG'
NGC6240 which has long been famous as the galaxy with the highest
luminosity in $\Hzwei$ emission \citep{jos84}.  Van der Werf et
al. (1993), \citet{tec00} and \cite{lut03} suggest that large scale
shocks are responsible for the strong $\Hzwei$ emission in
NGC~6240. Their line imaging indicates that the $\Hzwei$ emission in
NGC~6240 does not arise from the two nuclei proper but from an extended
region centered between these two nuclei. \citet{wer93}
determined an observed luminosity of $\sim$ $8\times 10^{7}$
L$_{\sun}$ for the $\Hzweiseins$ line. In F23578$-$5307, the observed
luminosity of the $\Hzweiseins$ line is $3.1\times 10^{8}~$L$_{\sun}$,
a factor 3.5 more than for NGC~6240, and well above the ULIRG IRAS
F09039$+$0503 ($1.1\times 10^{8}$ L$_{\sun}$, \citet{vei97b}), with
one of the strongest H$_{2}$ emission lines relative to Pa$\alpha$ in
the ULIRG population.  The equivalent width of 42.7\AA\ observed for
$\Hzweiseins$ in F23578$-$5307 is well above the $0-20$~\AA\ typical for
ULIRGs \citep{gol95} but short of the 71~\AA\ of NGC~6240. This confirms
that the large H$_2$ luminosity reflects an unusual type of activity
and is not just a consequence of scaling up a normal ULIRG spectrum.

Further studies of these systems may help determining the origin of
such extreme $\Hzwei$ luminosities. Is the situation for NGC6240,
i.e. strong emission between two nuclei of a merging system, typical?
Images of F23578$-$5307 taken at NTT/SOFI \citep[see][]{rig99}) and the
VLT/ISAAC (Tacconi, priv. communication) under good (0.8\arcsec)\/ and
excellent seeing (0.35\arcsec) conditions, respectively, show that
this ULIRG has a single nucleus. So far, there are no hints that this
extraordinary ULIRG has a similar morphology as NGC~6240 where the
projected seperation of the two nuclei is 0.75 kpcs. At the distance
of IRAS~F23578$-$5307 this would correspond to $\approx0.4$\arcsec,
allowing to barely resolve the two nuclei.

\section{Conclusions}
\label{Conclusions}
We have obtained moderate resolution near-infrared spectroscopy of a
sample of ULIRGs previously studied in the mid-infrared with ISO.  A
first goal was the search for near-infrared indications of AGN
activity, namely broad components to the hydrogen recombination lines
$\Pa$ or $\Pb$, and detection of the coronal line $\Si$ 1.962~$\mu$m.
Most spectra do not show these indicators and appear starburst-like.
In none of the 24 ULIRGs do we detect a broad velocity component of
the hydrogen recombination lines. In IRAS~04114E we have evidence for a
$\Si$ line with a luminosity of $1.4\times 10^{6}$ L$_{\sun}$ (no
extinction correction applied).  The agreement of near- and
mid-infrared classification is good, in line with previous suggestions
that AGN in ULIRGs are often visible over a wide wavelength range, and
with the majority of ULIRGs being predominantly starburst-powered.

We have revisited the `$\Brg$-deficit' in ULIRGs found by
\citet{gol95,gol97b}. Using the line pair $\Pa$, $\Ha$ for extinction
correction, we confirm their result by finding a $\Pa$-deficit.
However, when using the two near-infrared lines $\Pa$ and $\Brg$, the
inferred extinction is significantly higher and no deficit of ionizing
luminosity is found on average. This is consistent with most of the
luminosity in these sources being powered by star formation at
A$_V\sim10-20$ mag.

Many ULIRGs are found to be strong emitters of $\Fe$.  ULIRGs that are
optically classified as LINER or starburst also populate different
regions in a diagnostic diagram using the ratios
${\Fsechs}/{\Brg}$ and ${\Oi}/{\Ha}$. These observations are
consistent with LINERs in infrared galaxies being due to supernova and
superwind related shocks. Age effects and/or different location and
obscuration of the wind and starburst regions may contribute to making
a particular source a starburst or LINER type.

Two unusual objects were identified in the sample.  IRAS~F00183$-$7111
which ISO observation suggest to be an obscured AGN exhibits extremely
wide and strong lines of $\Fe$.  IRAS~F23578$-$5307 is one of
the most luminous infrared galaxies in $\Hzweiseins$ according to our knowledge.

\begin{acknowledgements}
       We like to thank Matthew D.  Lehnert for helpful discussion
during the data analysis and interpretation of the results and as well
Sylvain Veilleux for interesting conversation about part of the
results. Additionally, we are grateful to the staff of ESO for their
support with data acquisition and reduction. Finally, we would like to thank the anonymous referee for valuable comments. This research has made
use of the NASA/IPAC Extragalactic Database (NED) which is operated by
the Jet Propulsion Laboratory, California Institute of Technology,
under contract with the National Aeronautics and Space Administration.
\end{acknowledgements}
\appendix 
\section{K band Image of IRAS19458+0944}
\begin{figure}
\begin{center}
\includegraphics[angle=270,scale=0.32]{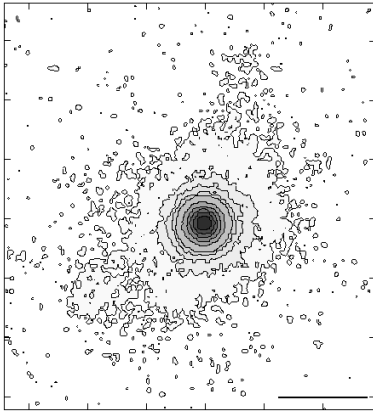}
\caption[]{\label{fig:k19458}K$_{s}$-band image of IRAS~19458 plotted
in grey scale and contour forms with a size of $\sim13\arcsec\times14\arcsec$. In an aperture of 4 arcseconds, the
magnitude in K$_{s}$ is 13.1~mag (Vega).  Angular scale on the sky is
indicated by the bar which represents 5~kpc. North is up, east to the
left.}
\end{center}
\end{figure}

The 5~minute K$_{s}$-band image taken at NTT/SOFI during the second
run (Fig.~\ref{fig:k19458}) shows a single system which seems to be
almost relaxed.  In an aperture of 4 arcseconds, the magnitude in
K$_{s}$ is 13.1~mag (Vega).
\bibliographystyle{aa} 

\end{document}